\PassOptionsToPackage{unicode}{hyperref}
\PassOptionsToPackage{hyphens}{url}
\PassOptionsToPackage{dvipsnames,svgnames,x11names}{xcolor}
\documentclass[
  12pt,
]{article}
\usepackage{amsmath,amssymb}
\usepackage{lmodern}
\usepackage{iftex}
\ifPDFTeX
  \usepackage[T1]{fontenc}
  \usepackage[utf8]{inputenc}
  \usepackage{textcomp} % provide euro and other symbols
\else % if luatex or xetex
  \usepackage{unicode-math}
  \defaultfontfeatures{Scale=MatchLowercase}
  \defaultfontfeatures[\rmfamily]{Ligatures=TeX,Scale=1}
\fi
\IfFileExists{upquote.sty}{\usepackage{upquote}}{}
\IfFileExists{microtype.sty}{% use microtype if available
  \usepackage[]{microtype}
  \UseMicrotypeSet[protrusion]{basicmath} % disable protrusion for tt fonts
}{}
\makeatletter
\@ifundefined{KOMAClassName}{% if non-KOMA class
  \IfFileExists{parskip.sty}{%
    \usepackage{parskip}
  }{% else
    \setlength{\parindent}{0pt}
    \setlength{\parskip}{6pt plus 2pt minus 1pt}}
}{% if KOMA class
  \KOMAoptions{parskip=half}}
\makeatother
\usepackage{xcolor}
\usepackage[margin=1in]{geometry}
\usepackage{graphicx}
\makeatletter
\def\maxwidth{\ifdim\Gin@nat@width>\linewidth\linewidth\else\Gin@nat@width\fi}
\def\maxheight{\ifdim\Gin@nat@height>\textheight\textheight\else\Gin@nat@height\fi}
\makeatother
\setkeys{Gin}{width=\maxwidth,height=\maxheight,keepaspectratio}
\makeatletter
\def\fps@figure{htbp}
\makeatother
\newlength{\cslhangindent}
\newlength{\csllabelwidth}
\newlength{\cslentryspacingunit} % times entry-spacing
\newenvironment{CSLReferences}[2] % #1 hanging-ident, #2 entry spacing
 {% don't indent paragraphs
  \setlength{\parindent}{0pt}
  \ifodd #1
  \let\oldpar\par
  \def\par{\hangindent=\cslhangindent\oldpar}
  \fi
  \setlength{\parskip}{#2\cslentryspacingunit}
 }%
 {}
\usepackage{calc}

\usepackage{xcolor}
\usepackage{soul}
\usepackage[margin=1in]{geometry} 
\usepackage[T1]{fontenc}
\usepackage{lmodern}
\usepackage{xurl}
\usepackage{longtable}
\graphicspath{ {images/} }
\usepackage{booktabs}
\usepackage{lscape}
\usepackage{dcolumn}
\usepackage{adjustbox}

\usepackage{setspace}

\usepackage[flushleft]{threeparttable}
\onehalfspace

\usepackage{hyperref}
\hypersetup{
    colorlinks=true,
    linkcolor=DarkBlue,
    filecolor=magenta,
    citecolor=Crimson,
    urlcolor=DarkBlue}

\usepackage{bm}
\usepackage{float}
\usepackage{multicol}
\usepackage{multirow}
\usepackage[utf8]{inputenc}
\usepackage{graphicx}
\usepackage{subfigure}
\usepackage{wrapfig}
\usepackage{afterpage}

\usepackage{tikz}

\usepackage[normalem]{ulem}
\usepackage{here}
\usepackage{amsmath,amsthm,amssymb}
\usepackage[all]{xy}
\usepackage{yhmath}

\DeclareMathOperator{\logit}{logit}

\ifdefined\knitrout
\renewenvironment{knitrout}{\begin{footnotesize}}{\end{footnotesize}}
\else
\fi

\makeatletter
\let\oldalltt\alltt
\def\alltt{\@ifnextchar[\alltt@i \alltt@ii}
\def\alltt@i[#1]{\oldalltt[#1]\footnotesize}
\def\alltt@ii{\oldalltt\footnotesize}
\makeatother

\definecolor{ForestGreen}{RGB}{34,139,34}
\definecolor{Crimson}{RGB}{220,20,60}
\definecolor{DarkRed}{RGB}{139,0,0}
\definecolor{DarkGreen}{RGB}{0,100,0}
\definecolor{DarkBlue}{RGB}{0,0,139}
\ifLuaTeX
  \usepackage{selnolig}  % disable illegal ligatures
\fi
\IfFileExists{bookmark.sty}{\usepackage{bookmark}}{\usepackage{hyperref}}
\IfFileExists{xurl.sty}{\usepackage{xurl}}{} % add URL line breaks if available
\hypersetup{
  pdftitle={Bayesian modeling of spatial ordinal data from health surveys},
  colorlinks=true,
  linkcolor={blue},
  filecolor={Maroon},
  citecolor={Blue},
  urlcolor={blue},
  pdfcreator={LaTeX via pandoc}}

\title{\huge Bayesian modeling of spatial ordinal data from health
surveys}
\author{\Large Beltrán-Sánchez, MA\(^1\); Martínez-Beneito, MA\(^2\);
Corberán-Vallet, A\(^3\)\\
\(^1\)\href{mailto:angel.beltran@uv.es}{\nolinkurl{angel.beltran@uv.es}};
\(^2\)\href{mailto:miguel.a.martinez@uv.es}{\nolinkurl{miguel.a.martinez@uv.es}};
\(^3\)\href{mailto:ana.corberan@uv.es}{\nolinkurl{ana.corberan@uv.es}}\\
Department of Statistics and Operations Research, University of
Valencia, Spain}
\date{August 2, 2023}

\begin{document}
\maketitle

\hypertarget{abstract}{%
\subsection*{Abstract}\label{abstract}}
\addcontentsline{toc}{subsection}{Abstract}

Health surveys allow exploring health indicators that are of great value
from a public health point of view and that cannot normally be studied
from regular health registries. These indicators are usually coded as
ordinal variables and may depend on covariates associated with
individuals. In this paper, we propose a Bayesian individual-level model
for small-area estimation of survey-based health indicators. A
categorical likelihood is used at the first level of the model hierarchy
to describe the ordinal data, and spatial dependence among small areas
is taken into account by using a conditional autoregressive (CAR)
distribution. Post-stratification of the results of the proposed
individual-level model allows extrapolating the results to any
administrative areal division, even for small areas. We apply this
methodology to the analysis of the Health Survey of the Region of
Valencia (Spain) of 2016 to describe the geographical distribution of a
self-perceived health indicator of interest in this region.

\textbf{Keywords:} Spatial statistics, ordinal data analysis,
survey-based studies, individual-level model, post-stratification

\hypertarget{sec:introduction}{%
\section{Introduction}\label{sec:introduction}}

Geographical studies in small areas are an excellent and widely used
epidemiological tool. They have been typically used to monitor health
problems based on data collected from disease registries or health
databases that contain observed counts of the event of interest
(e.g.~incident cases or death counts). A common characteristic of these
information sources is that they are exhaustive; that is, they collect
information on all health problems detected by health administrations.
The main objective of most of these studies is to derive reliable
statistical estimates of health outcomes for a set of small areas. This
can be achieved using Bayesian hierarchical spatial models that borrow
information between neighboring areas
(\protect\hyperlink{ref-Leroux2000}{Leroux \emph{et al.}, 2000};
\protect\hyperlink{ref-Lawson2008}{Lawson, 2008};
\protect\hyperlink{ref-Migue2019}{Martínez-Beneito \& Botella-Rocamora,
2019}).

Unfortunately, health registries only account for a fraction of the
information that would be of interest for epidemiological research. For
example, risk factors such as smoking, alcohol consumption or unhealthy
behaviors are rarely collected at health databases for the whole
population. It may also be of interest to analyze other health
indicators such as mental health issues, physical limitation problems,
social support needs or health habits with a potential impact on health.
These indicators can be obtained from health surveys, which are
collected periodically. Unfortunately, in this case, the information is
not available for the whole population, but just for a sample of
individuals. Survey data are collected following a specific sampling
scheme, for instance simple random sampling (SRS) or more complex
designs such as stratified or cluster sampling. Whatever the sampling
design, all the information used for the data collection mechanism
should be included in the analysis, or else inferences will not
necessarily be appropriate for the population of interest
(\protect\hyperlink{ref-Little2003}{Little, 2003};
\protect\hyperlink{ref-Gelman2013}{Gelman \emph{et al.}, 2013}).
Different estimates and modeling strategies have been proposed to
account for the sampling design. In a regression context, we can
distinguish between area-level and unit-level (individual-level) models,
as detailed below.

Survey-based small-area estimation has usually been carried out through
area-level model-based approaches, where the original individual-level
data are aggregated by area before being modeled. Within this approach,
non-spatial estimates are usually derived at a first stage, typically
taking into account the sampling design. This is the case of the
commonly used Horvitz-Thompson direct estimator, which computes a new
area-level outcome of interest weighting sampled individuals according
to their associated probability of being sampled
(\protect\hyperlink{ref-HT1952}{Horvitz \& Thompson, 1952}). In a second
stage, spatial dependence among these design-based estimates is then
considered by means of a spatial smoothing model that borrows
information between areas (\protect\hyperlink{ref-You2011}{You \& Zhou,
2011}; \protect\hyperlink{ref-Marhuenda2013}{Marhuenda \emph{et al.},
2013}; \protect\hyperlink{ref-Chen2014}{Chen \emph{et al.}, 2014};
\protect\hyperlink{ref-Mercer2014}{Mercer \emph{et al.}, 2014};
\protect\hyperlink{ref-Porter2014}{Porter \emph{et al.}, 2014};
\protect\hyperlink{ref-Paige2020}{Paige \emph{et al.}, 2020}). These two
separate phases produce indirect spatial estimates accounting for the
sampling design; nevertheless, the effect of considering two separate
estimation phases is not well-known and may introduce biases by
simplifying the nature of the problem
(\protect\hyperlink{ref-Desmee2015}{Desmée \emph{et al.}, 2015};
\protect\hyperlink{ref-Mauff2020}{Mauff \emph{et al.}, 2020}). Indeed,
the direct area-level estimates produced in the first step are used to
feed the second stage without a clear idea of the effect of the
simplifying assumptions done in each of these steps. In addition, most
of these works consider Normal likelihood models to describe the
aggregated data (\protect\hyperlink{ref-FH1979}{Fay \& Herriot, 1979}).
However, when small-area sample sizes are small or even zero, the
Central Limit Theorem does not hold, so this assumption cannot be
reasonably assumed for very small areas (say with less than a few tens
of sampled units per area). Moreover, in small-area studies, Normal
likelihood models should not be considered if the new area-level
response variable is restricted to an interval. For example, if the
original individual-level variable is binary, an area-level quantity of
interest could be the proportion of successes in each area. For these
proportions, which are restricted to the interval \([0,1]\), normality
should not be assumed.

Individual-level spatial models have also been proposed to allow
estimates of the effect of any covariate associated with the outcome of
interest (\protect\hyperlink{ref-Mercer2015}{Mercer \emph{et al.},
2015}; \protect\hyperlink{ref-Vandendijck2016}{Vandendijck \emph{et
al.}, 2016}; \protect\hyperlink{ref-Watjou2017}{Watjou \emph{et al.},
2017}; \protect\hyperlink{ref-Parker2020}{Parker \emph{et al.}, 2020},
\protect\hyperlink{ref-Parker2022}{2022};
\protect\hyperlink{ref-Congdon2022}{Congdon \& Lloyd, 2022};
\protect\hyperlink{ref-Sun2022}{Sun \emph{et al.}, 2022};
\protect\hyperlink{ref-Thompson2022}{Thompson \emph{et al.}, 2022};
\protect\hyperlink{ref-Vergara2023}{Vergara-Hernández \emph{et al.},
2023}). As previously mentioned, all the information related to the data
collection mechanism should be included in the model. This is typically
done by modeling the response variable conditional on the design
variables, if available, or by including the design weights in the
model. Different likelihoods can be assumed depending on the response
variable, and estimates or predictions can be made at either the
individual level or aggregated up to any desired level of the hierarchy
via post-stratification (\protect\hyperlink{ref-Little1993}{Little,
1993}). This statistical procedure is known in the literature as
multilevel regression with post-stratification (MRP)
(\protect\hyperlink{ref-Gelman1997}{Gelman \& Little, 1997};
\protect\hyperlink{ref-Park2004}{Park \emph{et al.}, 2004},
\protect\hyperlink{ref-Park2006}{2006}). Recently, a correlated spatial
random effect has been included in a model to explain spatial dependence
between neighboring areas when the outcome of interest is binary
(\protect\hyperlink{ref-Vergara2023}{Vergara-Hernández \emph{et al.},
2023}). Once the model has been fitted, individual estimates are
post-stratified to reproduce these effects on the geographical areas of
interest according to their population totals, adjusting for selection
bias and correcting for differences between the sample composition and
that of the geographical areas of interest.

Usually, health survey questions have different response options, so
that respondents answer the one that seems most appropriate to them. In
addition, these answer options are often ordered or ranked. For
instance, a question about smoking frequency may have four response
levels: \(1\) \(=\) Never smoked/only tried; \(2\) \(=\) Don't smoke but
have smoked; \(3\) \(=\) Smoke but not daily; and \(4\) \(=\) Smoke
daily. Therefore, it is common to find ordinal data in survey-based
studies. However, to our knowledge, there is currently no literature on
spatial models for ordinal survey-based data. The analysis of ordinal
data is a non-standard statistical area. In fact, the most typical
likelihood families do not fit to data of this kind. As a consequence of
this complexity, spatial dependence is usually ignored, although survey
respondents have spatial attributes. The spatial analysis of ordinal
survey-based data merges several statistical research branches, at least
spatial statistics, analysis of ordinal data and analysis of survey data
with complex sampling designs, making the development of methods for
data of that kind a particularly challenging task.

The main objective of this paper is to develop a Bayesian hierarchical
individual-level model for small-area analyses from ordinal survey-based
data. The model, which simultaneously accounts for spatial dependence
and sampling design information at a single stage, allows estimating
finite quantities of the population of interest under a known
informative sampling design via the MRP approach. Hence, it entails an
expansion of applications in spatial small-area epidemiological studies
to the spatial analysis of ordinal survey-based data. By using this
methodology, ordinal data from health surveys can be easily summarized
and exploited to a greater extent, taking into account the spatial
nature of the data.

The paper is organized as follows.
\protect\hyperlink{sec:methodology}{Section 2} reviews individual-level
models for small-area survey-based studies within the MRP approach. Our
proposed model for ordinal data within this context is subsequently
introduced. \protect\hyperlink{sec:case-study}{Section 3} applies this
methodology to the Health Survey of the Region of Valencia (Spain) of
2016 (HSRV2016). Finally, \protect\hyperlink{sec:conclusions}{Section 4}
concludes with some final remarks and future work.

\hypertarget{sec:methodology}{%
\section{Methodology}\label{sec:methodology}}

In this section, we first provide background on individual-level models
for small-area survey-based studies within the MRP approach. Next, we
detail a proposal for ordinal data analysis within this framework.

\hypertarget{sec:background}{%
\subsection{Background}\label{sec:background}}

Survey-based data analyses should include all the information used for
the sampling design, otherwise model-based estimates will be subject to
potentially large biases (\protect\hyperlink{ref-Little2003}{Little,
2003}; \protect\hyperlink{ref-Gelman2013}{Gelman \emph{et al.}, 2013}).
This can be done by modeling the individual-level response variable
conditional on the design variables. For instance, if a cluster sampling
scheme is followed, a cluster-level random effect may be used in the
model to take this into account. If a stratified design is used instead,
one should include fixed effects accounting for the strata. The idea is
that when all the design variables are taken into account in the model,
the conditional distribution of the response variable given the
covariates for the sampled units is independent of the inclusion
probabilities, and so we can consider the sampling design as ignorable.
If the design variables were unknown, the sampling design could be
alternatively taken into account by including the design weights in the
linear predictor (\protect\hyperlink{ref-Vandendijck2016}{Vandendijck
\emph{et al.}, 2016}; \protect\hyperlink{ref-Watjou2017}{Watjou \emph{et
al.}, 2017}) or by considering a pseudo-likelihood approach
(\protect\hyperlink{ref-Binder1983}{Binder, 1983};
\protect\hyperlink{ref-Skinner1989}{Skinner, 1989}). From now on, we
assume a context where all the design variables are known.

Usually, individual-level models incorporate demographic and/or
geographic subgroups as covariates (e.g.~sex, age group, ethnicity,
education, geographic location, etc.). The reason is that either these
categorical covariates are involved in the sampling scheme, such as the
strata cells in a stratified sampling design, or simply they have a
relevant effect on the response variable. Whatever the reason, the
combination of all the categories of these covariates defines the
so-called post-stratification cells, that will be later on used to
derive population estimates at any level of the hierarchy, via
post-stratification, correcting any potential difference between the
sample and population compositions. Within the MRP approach,
ignorability corresponds to the assumption of SRS within each
post-stratification cell; that is, equal inclusion probabilities for the
individuals within cells (\protect\hyperlink{ref-Gelman2007}{Gelman,
2007}). If this assumption is met, unbiased estimates of the effect of
these cells are easily obtained and post-stratification takes care of
correcting any potential discrepancy between the sample and the
population corresponding to each cell, thus producing adequate estimates
at the area/group level of interest. Different survey-based studies have
used the MRP approach to derive weighted survey estimates
(\protect\hyperlink{ref-Zhang2018}{Zhang \emph{et al.}, 2014};
\protect\hyperlink{ref-Wang2015}{Wang \emph{et al.}, 2015};
\protect\hyperlink{ref-Buttice2017}{Buttice \& Highton, 2017};
\protect\hyperlink{ref-Downes2018}{Downes \emph{et al.}, 2018};
\protect\hyperlink{ref-Gelman2018}{Gelman \emph{et al.}, 2018};
\protect\hyperlink{ref-Downes2020}{Downes \& Carlin, 2020};
\protect\hyperlink{ref-Gao2021}{Gao \emph{et al.}, 2021}). These
studies, which traditionally use just independent random effects, do not
account for spatially structured variability.

Within the context of the MRP approach, Vergara-Hernández \emph{et al.}
(\protect\hyperlink{ref-Vergara2023}{2023}) have recently proposed a
novel approach to estimate the proportions of a binary variable within
the small-area estimation framework. In that work, the authors present a
logistic individual-level model that integrates the sampling scheme
followed (e.g.~stratified sampling), auxiliary variables and
geographical variability. That approach can be summarized as follows:
Let \(Y_i \in \{0,1\}\) be a binary response variable for the \(i\)-th
individual of the population of interest (\(i = 1,\dots,N\)). Let us
assume that the study region is divided in \(K\) small areas, being
\(\{N_k\}_{k=1}^K\) their known population sizes, so that
\(N = \sum_{k=1}^K N_k\) stands for the total population size of the
study region. A sample of size \(n_k\) is drawn from each area \(k\)
following a known sampling scheme (e.g.~SRS, stratified or cluster
sampling), where some of the \(n_k\) could be zero (e.g.~small areas not
sampled within a cluster sampling design). Hence, the total sample size
is \(n = \sum_{k = 1}^K n_k < N\), which means that only some values of
\(Y_i\) are observed, the ones corresponding to the \(n\) survey
respondents. Without loss of generality, we assume that the first \(n\)
individuals in the population correspond to the survey respondents,
while the following are all unobserved; that is, the observed values are
\(y_i\), \(i = 1,\dots,n\). Let \(\boldsymbol{x}_i = (z_i, m_i)^{T}\)
denote the vector containing the value of a categorical covariate
\((z_i \in \{1, \dots, H\})\) and the small area
\((m_i \in \{1,\dots,K\})\), both corresponding to the \(i\)-th
respondent. Note that \(z_i\) could be either a stratification variable
used for the sampling design or an auxiliary variable used to improve
the proportion estimates. If more than one variable of this kind were
available, they could be considered accordingly in the subsequent model.

The following likelihood is assumed: \begin{equation*}
y_i|\pi(\boldsymbol{x}_i) \sim Bernoulli(\pi(\boldsymbol{x}_i)),
\end{equation*} where
\(\pi(\boldsymbol{x}_i) = P(Y_i = 1 | \boldsymbol{x}_i)\) is modeled as:
\begin{equation}
\logit (\pi(\boldsymbol{x}_i)) = \mu + \alpha_{z_i} + \theta_{m_i}. \label{eq:backmodel}
\end{equation} If \(z_i\) was the only variable involved in the sampling
scheme, given \(\boldsymbol{x}_i\), the individuals would be sampled
following SRS within each level and, therefore, the sampling design
would be ignorable. In this proposal, the intercept is assumed to have
an improper uniform prior; that is, \(p(\mu)\propto 1\); the effect of
the auxiliary (categorical) covariate is modeled such that \(\alpha_1\)
is set to \(0\), while improper uniform priors are also considered for
\(\alpha_2, \dots, \alpha_H\); and a spatial CAR prior
(\protect\hyperlink{ref-Leroux2000}{Leroux \emph{et al.}, 2000}) is used
to induce spatial dependence on
\(\boldsymbol{\theta} = (\theta_1,\dots,\theta_K)^{T}\).

Note that we have denoted the probability of success for each individual
as \(\pi(\boldsymbol{x}_i)\) in order to be consistent with the
individual-level notation. However, these probabilities can only take as
many values as combinations of the \(H\) levels of the categorical
covariate and the \(K\) small areas; that is, respondents with the same
value of \(z_i\) and \(m_i\) (equivalently \(\boldsymbol{x}_i\)) have
the same value of \(\pi(\boldsymbol{x}_i)\) according to Model
\eqref{eq:backmodel}. Therefore, given the vector
\(\boldsymbol{x} = (h, k)^{T}\), we can alternatively denote those
probabilities as simply \(\pi_{hk}\). Once the model has been fitted,
the derived \(\hat{\pi}_{hk}\) estimates can be subsequently
post-stratified to compute population estimates at a small-area level.
For instance, the real proportion of success for the \(k\)-th areal unit
can be estimated post-stratifying the cell estimates \(\hat{\pi}_{hk}\)
as (\protect\hyperlink{ref-Gelman1997}{Gelman \& Little, 1997};
\protect\hyperlink{ref-Park2004}{Park \emph{et al.}, 2004},
\protect\hyperlink{ref-Park2006}{2006}): \begin{equation*} 
\hat{P}_k = \frac{1}{N_k} \sum_{h = 1}^H N_{hk} \hat{\pi}_{hk},
\end{equation*} where \(N_{hk}\) is the population size of level \(h\)
and area \(k\). This post-stratification process converts the model
estimates into population quantities that incorporate the composition of
each area in terms of the auxiliary/stratifying variable. In summary,
that proposed methodology merges tools accounting for sampling designs,
the use of auxiliary variables and spatial methods in a common framework
in order to derive sensible small-area estimates taking the data
collection process into account.

\hypertarget{sec:ordinal}{%
\subsection{Our proposal: A regression model for small-area inference
with ordinal data}\label{sec:ordinal}}

Following the guidelines described previously, which allow us to take
into account both the sampling design and spatial dependence of the
data, we develop here a methodology for the analysis of spatial ordinal
survey-based data. To our knowledge, this has not been previously
proposed in the spatial literature.

Let \(Y_i \in \{1,\dots,J\}\) be an ordinal response variable that
quantifies a health indicator for the \(i\)-th individual of the
population of interest, \(i = 1,\dots,N\). Let
\(\boldsymbol{y} = (y_1,\dots,y_n)\) denote the vector containing the
observed values for all respondents. Let us now assume that
\(N_{(\cdot)}\) and \(n_{(\cdot)}\) represent, respectively, the
population and sample size according to their subscript. Thus,
\(\{N_k\}_{k = 1}^K\) and \(\{n_k\}_{k = 1}^K\) have exactly the same
meaning as before. First, we will focus on the case of a single
categorical covariate \(z_i\) with \(H\) levels, as for example the
stratum in a stratified sampling design. In that case,
\(\boldsymbol{x}_i = (z_i, m_i)^{T}\) is the covariate vector that
indicates the level \(h\) (\(z_i = h \in \{1, \dots, H\}\)) and the
small area \(k\) (\(m_i = k \in \{1, \dots, K\}\)) to which the \(i\)-th
respondent belongs.

The following likelihood is assumed: \begin{equation}
y_i|\boldsymbol{\pi}(\boldsymbol{x}_i) \sim Categorical(\boldsymbol{\pi}(\boldsymbol{x}_i)), \label{eq:likelihood}
\end{equation} where
\(\boldsymbol{\pi}(\boldsymbol{x}_i) = (\pi_1(\boldsymbol{x}_i), \pi_2(\boldsymbol{x}_i), \dots, \pi_J(\boldsymbol{x}_i))^{T}\)
is the vector that contains the probabilities for the different
categories associated with the \(i\)-th respondent given
\(\boldsymbol{x}_i\); that is,
\(\pi_j(\boldsymbol{x}_i) = P(Y_i = j | \boldsymbol{x}_i)\),
\(j = 1, \dots, J\). Since \(Y_i\) is ordinal, it seems more natural to
model its cumulative probabilities instead of the probability of each
category (\protect\hyperlink{ref-McCull1980}{McCullagh, 1980};
\protect\hyperlink{ref-Congdon2005}{Congdon, 2005}). Namely, let
\(\gamma_j(\boldsymbol{x}_i)\) denote the cumulative probability of
categories \(1\) to \(j\) given \(\boldsymbol{x}_i\); that is:
\begin{equation*} 
\gamma_j(\boldsymbol{x}_i) = P(Y_i \leq j | \boldsymbol{x}_i) = \pi_1(\boldsymbol{x}_i) + \pi_2(\boldsymbol{x}_i) + \dots + \pi_j(\boldsymbol{x}_i), \ j = 1, \dots, J,
\end{equation*} where
\(0 < \gamma_1(\boldsymbol{x}_i) < \gamma_2(\boldsymbol{x}_i) < \dots < \gamma_{J-1}(\boldsymbol{x}_i) < \gamma_{J}(\boldsymbol{x}_i) = 1\).
In that case, we model the first \(J-1\) cumulative probabilities as:
\begin{equation}
\logit (\gamma_j(\boldsymbol{x}_i)) = \kappa_j + \alpha_{z_i} + \theta_{m_i}, \ j = 1,\dots,J-1, \label{eq:model}
\end{equation} where \(\kappa_1 < \kappa_2 < \dots < \kappa_{J-1}\).
This proposal falls within the Proportional Odds Logistic Regression
(POLR) models (\protect\hyperlink{ref-McCull1980}{McCullagh, 1980}).

\begin{figure}[!h]
\centering
\includegraphics[width = 16.5cm]{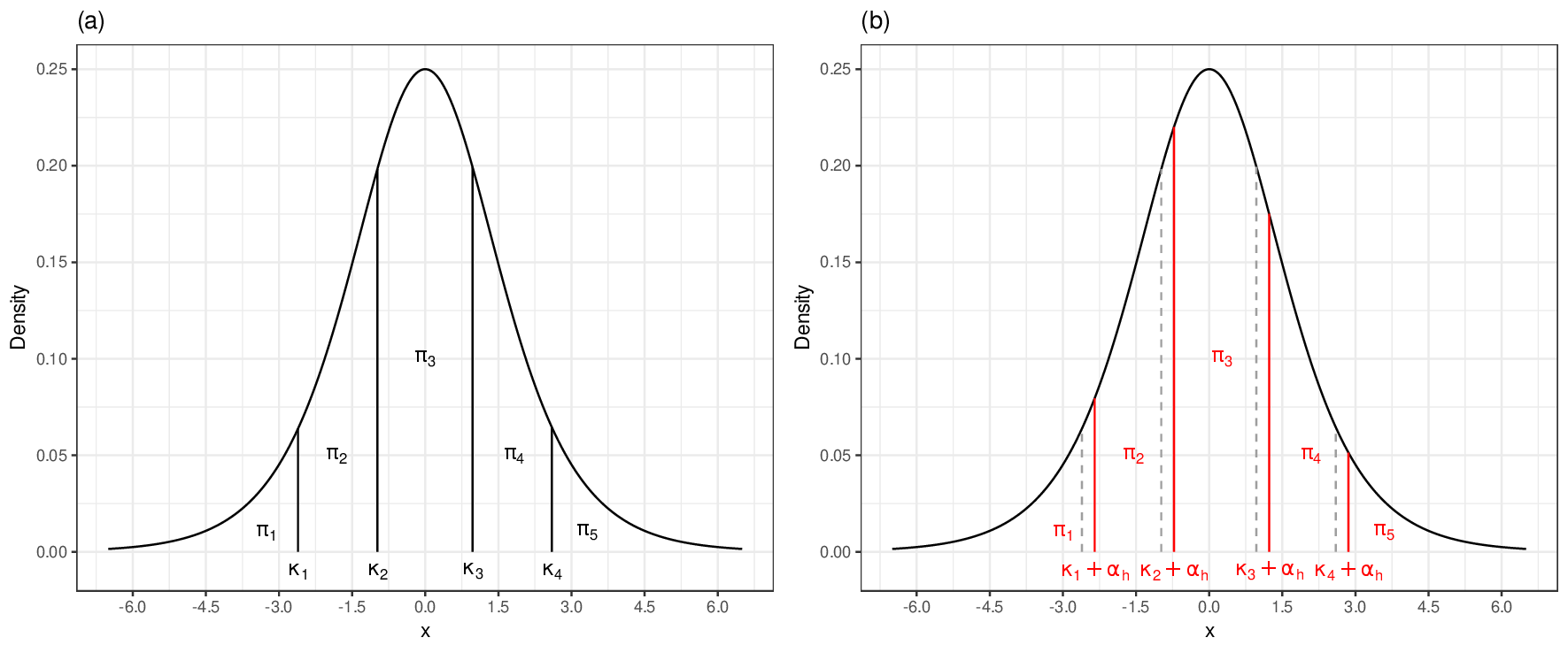}
\caption{(a) An illustrated example of the interpretation of $\{\kappa_j\}_{j = 1}^{J-1}$ from Model \eqref{eq:model} as cut points of the standard logistic probability density function: $f(x) = \exp\{x\}/(1+\exp\{x\})^2$. (b) A positive value of $\alpha_{z_i}$ from Model \eqref{eq:model} would move all cut points towards the right a distance equal to the value of that effect, changing the relative probabilities of all five categories.}
\label{fig:cutpoints}
\end{figure}

Note that if \(\alpha_{z_i}\) and \(\theta_{m_i}\) (covariate effects)
were removed from Model \eqref{eq:model}, all individuals would share
the same vector of probabilities, independently of their stratum and
small area. In that case,
\(\gamma_j(\boldsymbol{x}_i) = \logit^{-1}(\kappa_j) = \exp\{\kappa_j\}/(1 + \exp\{\kappa_j\}) = F(\kappa_j)\),
\(j=1,\dots,J-1\), where \(F\) denotes the standard logistic cumulative
distribution function. So, the set of parameters
\(\kappa_1 < \kappa_2 < \dots < \kappa_{J-1}\) could be interpreted as
cut points of the standard logistic probability density function,
establishing the probability of each category as the area under that
density function between the two corresponding consecutive cut points
(\protect\hyperlink{ref-Congdon2005}{Congdon, 2005};
\protect\hyperlink{ref-Faraway2016}{Faraway, 2016};
\protect\hyperlink{ref-Dobson2018}{Dobson \& Barnett, 2018}). This is
illustrated in Figure \ref{fig:cutpoints} (a) for an arbitrarily chosen
set of values. In addition, positive values of \(\alpha_{z_i}\)
(equivalently \(\theta_{m_i}\)) increase the values of
\(\gamma_j(\boldsymbol{x}_i)\) for the corresponding level (or the
corresponding small area), and so low values of \(Y_i\) are more likely
to occur. In particular, the probability of the first category of
\(Y_i\) increases, while the last one decreases. Negative values of
\(\alpha_{z_i}\) or \(\theta_{m_i}\) have the opposite effect. This is
illustrated in Figure \ref{fig:cutpoints} (b) for the case of an
arbitrary positive effect \(\alpha_h\). Finally, in the complete Model
\eqref{eq:model},
\(\gamma_j(\boldsymbol{x}_i) = \logit^{-1}(\kappa_j + \alpha_{z_i} + \theta_{m_i}) = F(\kappa_j + \alpha_{z_i} + \theta_{m_i})\),
\(j=1,\dots,J-1\). So, the set of parameters
\((\kappa_j + \alpha_{z_i} + \theta_{m_i})_{j = 1}^{J-1}\) are the cut
points of the standard logistic probability density function for the
stratum and small area corresponding to the \(i\)-th respondent.

\hypertarget{sec:priors}{%
\subsubsection{Prior distributions}\label{sec:priors}}

We assume \(\alpha_1 = 0\) and \(\{p(\alpha_h)\}_{h=2}^H \propto 1\);
that is, improper uniform priors for the fixed effects associated with
the categorical covariate. Instead of assuming a corner constraint as
this, we could alternatively assume a zero-sum constraint; that is,
\(\alpha_1 = -\sum_{h = 2}^H \alpha_h\), depending on the convenience of
parameterization of the model. Regarding the spatial random effect, we
assume \(\boldsymbol{\theta}\) to follow a Leroux CAR prior distribution
(\protect\hyperlink{ref-Leroux2000}{Leroux \emph{et al.}, 2000}); that
is,
\((\boldsymbol{\theta}|\sigma^2, \lambda) \sim LCAR(\sigma^2, \lambda)\).
For the hyperparameters of the spatial random effect, we assume
non-informative uniform prior distributions: \(\sigma \sim U(0, 10)\)
and \(\lambda \sim U(0, 1)\). The upper limit in the first of these
uniform distributions is intended to assume a vague prior distribution
on \(\sigma\). In addition, we assume a total of \(H\) zero-sum
constraints for the spatial random effect, one for each level of the
categorical covariate. This is done in order to avoid spatial
confounding problems between fixed and spatial random effects, that
could have some influence on the post-stratification phase of the model,
to be introduced below (\protect\hyperlink{ref-Reich2006}{Reich \emph{et
al.}, 2006}; \protect\hyperlink{ref-Hodges2010}{Hodges \& Reich, 2010};
\protect\hyperlink{ref-Mari2014}{Marí-Dell'Olmo \emph{et al.}, 2014};
\protect\hyperlink{ref-Migue2019}{Martínez-Beneito \& Botella-Rocamora,
2019}). Specifically, the constraints are formulated as zero-sum
constraints for the spatial effects associated with respondents that
fall within the same level \(h\); that is,
\(\sum_{i: z_i = h} \theta_{m_i} = 0\), \(h=1,\dots,H\). In fact, those
constraints are equivalent to assuming orthogonality restrictions
between the vector \((\theta_{m_i})_{i = 1}^n\) and each column of the
design matrix of fixed effects
\((z_{ih})_{1 \leq i \leq n, 1 \leq h \leq H}\), where \(z_{ih} = 1\) if
\(z_i = h\) and \(0\) otherwise. Note that these orthogonality
restrictions can be seen simply as weighted constraints by the number of
respondents in each small area within each level \(h\); that is,
\(\sum_{k = 1}^K n_{hk}\theta_k = 0\), \(h=1,\dots,H\).

The prior distribution for vector
\(\boldsymbol{\kappa} = (\kappa_1,\dots,\kappa_{J-1})^{T}\) is defined
according to two premises: (i) all the categories should have equal
prior probabilities and (ii) it should fulfill the order constraint
\(\kappa_1 < \kappa_2 < \dots < \kappa_{J-1}\). This can be achieved by
including in the model a new vector
\(\boldsymbol{\delta} = (\delta_1,\dots,\delta_J)^{T}\), where
\(0 < \delta_j < 1\), \(j = 1,\dots,J\), and
\(\sum_{j = 1}^J \delta_j = 1\), so that
\(\boldsymbol{\delta} \sim Dirichlet(\boldsymbol{1}_J)\). Then, if we
define the vector \(\boldsymbol{\kappa}\) as a monotonic transformation,
into the whole real line, of the cumulative sums of the components of
\(\boldsymbol{\delta}\), assumptions (i) and (ii) are met. Specifically,
\(\kappa_j\) is defined as: \begin{equation}
\kappa_j = \logit \left(\sum_{r = 1}^j \delta_r\right), \ j = 1,\dots,J-1, \label{eq:cutpoints}
\end{equation} which takes the values of \(\boldsymbol{\delta}\),
defined in the \((0,1)\) interval, to the whole real line and satisfies
the order constraint of the components of \(\boldsymbol{\kappa}\).
However, direct modeling of the vector \(\boldsymbol{\delta}\) with a
Dirichlet prior distribution is not possible to implement in
\texttt{WinBUGS}, the software used in our case study. Alternatively,
let us assume that \(\boldsymbol{\delta}\) follows a stick-breaking
process (\protect\hyperlink{ref-Sethuraman1994}{Sethuraman, 1994}). The
process is based on the idea of breaking a ``stick'' of unit length into
\(J\) pieces similar to each other. To do this, a different proportion
of stick is taken in each iteration depending on the length of the stick
remaining in the previous one. That is, let
\(\boldsymbol{\omega} = (\omega_1,\dots,\omega_{J-1})^{T}\) be the
vector containing the successive proportions of stick taken in each
iteration. The stick-breaking process can be defined as follows:
\begin{equation*}
\omega_j \sim Beta(1, J-j), \ j=1,\dots,J-1.
\end{equation*} Defining the lengths of the pieces as: \begin{align*}
\delta_1 &= \omega_1, \\
\delta_j &= \prod_{r=1}^{j-1} (1-\omega_r)\omega_j, \ j=2,\dots,J-1, \\
\delta_J &= \prod_{r=1}^{J-1} (1-\omega_r),
\end{align*} we ensure that the sum of the lengths equals 1; that is,
\(\sum_{j = 1}^J \delta_j = 1\). That stick-breaking process is
equivalent to assuming a Dirichlet prior distribution on
\(\boldsymbol{\delta}\) with its vector of parameters equal to a vector
of ones (\protect\hyperlink{ref-Sethuraman1994}{Sethuraman, 1994};
\protect\hyperlink{ref-Paisley2010}{Paisley, 2010}). Hence, as wanted,
the stick-breaking process allows the \(J\) categories to have equal
prior probabilities. Note that we assume the \(\logit\) function to
transform the vector \(\boldsymbol{\delta}\) into
\(\boldsymbol{\kappa}\) (see Expression \eqref{eq:cutpoints}) due to the
interpretation of the components of \(\boldsymbol{\delta}\). In fact, if
the covariate effects were removed from Model \eqref{eq:model}, each
cumulative probability would be equal to the corresponding cumulative
sum of the components of \(\boldsymbol{\delta}\); that is, the
components of \(\boldsymbol{\delta}\) could be interpreted as the
probabilities for the different categories.

\hypertarget{sec:extension}{%
\subsubsection{Extending the model}\label{sec:extension}}

Note that the fixed effects \(\{\alpha_h\}_{h=2}^H\) in Model
\eqref{eq:model} have, in the logit scale, the same effect across the
different categories of the ordinal response variable, which could be a
strong assumption. In order to alleviate this, it is possible to allow
the covariate of interest to have different effects for each level of
the cumulative probabilities and, consequently, for each level of the
response variable. This extension of the model, which falls within
Partial Proportional Odds Logistic Regression (PPOLR) models
(\protect\hyperlink{ref-Peterson1990}{Peterson \& Harrell, 1990}), has
the form: \begin{equation}
\logit (\gamma_j(\boldsymbol{x}_i)) = \kappa_j + \alpha_{z_i, j} + \theta_{m_i}, \ j = 1,\dots,J-1, \label{eq:model2}
\end{equation} but it does not guarantee the order constraint of the
cumulative probabilities \(\gamma_j(\boldsymbol{x}_i)\). Alternatively,
Model \eqref{eq:model2} is equivalent to assuming that the intercepts
\(\kappa_j\) interact with the covariate of interest. Therefore, they
vary for each \(j\) and each value of the categorical covariate:
\begin{equation}
\logit (\gamma_j(\boldsymbol{x}_i)) = \kappa_{z_i, j} + \theta_{m_i}, \ j = 1,\dots,J-1. \label{eq:model3}
\end{equation} Model \eqref{eq:model3} ensures the order constraint
\(0 < \gamma_1(\boldsymbol{x}_i) < \gamma_2(\boldsymbol{x}_i) < \dots < \gamma_{J-1}(\boldsymbol{x}_i) < 1\),
for every value of \(z_i\), by means of a collection of stick-breaking
processes as prior distributions for the rows of the matrix
\(\boldsymbol{\kappa} = (\kappa_{hj})_{1 \leq h \leq H, 1 \leq j \leq J-1}\).

It should be noted that Model \eqref{eq:model3} becomes more complex if
one introduces more than one categorical covariate interacting with
\(\kappa_j\) to explain the ordinal response variable. In fact, in
complex sampling designs, it is common to have different combinations of
strata based on demographic-geographic characteristics, such as sex, age
group, ethnicity, education, geographic location, etc. In that
situation, the term \(\kappa_{z_i, j}\) in Model \eqref{eq:model3}
should be changed by \(\kappa_{z_i,w_i,\dots,j}\) where the subindex
depends on all the covariates interacting with \(\kappa_j\).

\hypertarget{sec:post-stratification}{%
\subsubsection{Post-stratification}\label{sec:post-stratification}}

Once the model has been fitted, we obtain as a result the estimates of
the probabilities of the different categories for all respondents; that
is, \(\hat{\pi}_j(\boldsymbol{x}_i)\), \(j = 1,\dots,J\),
\(i = 1,\dots,n\). These estimates correspond to each combination of the
\(H\) levels of the categorical covariate and \(K\) small areas. If
small-area estimates are pursued, these individual estimates should be
post-stratified in order to reproduce the strata composition and
particular effect of each geographical location. Usually, there is an
interest in estimating finite population quantities, such as the
proportion of individuals in each category of the response variable in
area \(k\): \begin{equation*}
P_{jk} = \frac{1}{N_k}\sum_{i:m_i = k} [Y_i = j] = \frac{1}{N_k}\sum_{h = 1}^H \sum_{\substack{i:m_i = k \\ z_i = h}} [Y_i = j], \ j=1,\dots,J,
\end{equation*} where \([Y_i = j]\) takes the value 1 if \(Y_i = j\) and
\(0\) otherwise. The proportions \(P_{jk}\), \(j = 1,\dots,J\) for a
particular small area \(k\) could be estimated as: \begin{equation*}
\hat{P}_{jk} = \frac{1}{N_k} \sum_{h = 1}^H\left(\sum_{i \in s_{hk}} [y_i = j] + \sum_{i \in s_{hk}'} \widehat{[Y_i = j]}\right), \ j=1,\dots,J,
\end{equation*} where \(s_{hk}\) and \(s_{hk}'\) denote the sets of
sampled and non-sampled individuals from level \(h\) and area \(k\),
respectively. For simplicity, we assume throughout this paper that
\(n_{hk}/N_{k}\) is small for all \(h\) and \(k\), which is a common
scenario in survey studies. So, finite population quantities of interest
are essentially the same as the corresponding superpopulation quantities
(\protect\hyperlink{ref-Gelman2007}{Gelman, 2007};
\protect\hyperlink{ref-Vergara2023}{Vergara-Hernández \emph{et al.},
2023}). Therefore, we can ignore in practice all finite population
corrections and the real proportions \(P_{jk}\), \(j = 1,\dots,J\) could
be simply estimated following the MRP approach as
(\protect\hyperlink{ref-Gelman1997}{Gelman \& Little, 1997};
\protect\hyperlink{ref-Park2004}{Park \emph{et al.}, 2004},
\protect\hyperlink{ref-Park2006}{2006}): \begin{equation}
\hat{P}_{jk} = \frac{1}{N_k} \sum_{h = 1}^H N_{hk} \hat{\pi}_{jhk}, \ j=1,\dots,J, \label{eq:post}
\end{equation} where \(\hat{\pi}_{jhk}\) is the estimated probability of
category \(j\) in level \(h\) and area \(k\). Indeed, Expression
\eqref{eq:post} yields weighted survey estimates that correct for
potential differences between the composition of the samples and those
of the respective populations. Note that this estimate requires knowing
the population size of each level \(h\) and area \(k\), \(N_{hk}\),
which will be known in general if the categorical covariate is a
stratifying factor of the sampling design or an auxiliary variable
(\protect\hyperlink{ref-Vergara2023}{Vergara-Hernández \emph{et al.},
2023}).

\hypertarget{sec:case-study}{%
\section{A case study}\label{sec:case-study}}

In this section, we apply the methodology developed in
\protect\hyperlink{sec:ordinal}{Section 2.2} to the study of
self-perceived health, one of the variables included in the HSRV2016.
This variable, which can be interpreted as an assessment of
self-perceived health status, which is a good general summary of health
status, was recorded for the population over \(14\) years old. The
variable has \(J = 5\) response levels: \(1\) \(=\) Very good; \(2\)
\(=\) Good; \(3\) \(=\) Regular; \(4\) \(=\) Bad; and \(5\) \(=\) Very
bad, ordered from better to worse self-perceived health.

The sampling design of the HSRV2016 is described below in order to
incorporate it into the statistical analysis. A total of \(5485\)
surveys were conducted following a two-stage sampling scheme. The first
sampling stage corresponds to a stratified sampling design, where the
sample units are the dwellings and the strata are formed according to
the age composition of the members of each dwelling. Specifically, there
are \(4\) strata defined as: \(1\) \(=\) Dwellings with 1 minor (without
residents over 74); \(2\) \(=\) Dwellings with 2 or more minors (without
residents over 74); \(3\) \(=\) Dwellings with residents over 74; and
\(4\) \(=\) Other dwellings. For each of the \(24\) health departments
in the Region of Valencia (RV) a stratified sample with the mentioned
strata was drawn and simple random sampling within each stratum. In the
second stage, the sample units are the residents of the dwellings
selected in the previous stage. In particular, surveys are collected
from all residents over \(64\) years old, and \(50\%\) of residents
between \(15\) and \(64\) by SRS.

To obtain small-area estimates, with considerable geographical
disaggregation, we consider here the municipality as the spatial unit of
interest, with a total of \(542\) municipalities in the RV. Let
\(\boldsymbol{x}_i = (s_i, a_i, d_i, m_i)^{T}\) be the vector containing
the values of the categorical covariates and small area (municipality)
for the \(i\)-th respondent, where:

\begin{itemize}
\item
  \(s_i \in \{1, 2\}\) denotes the sex: \(1\) \(=\) Male and \(2\) \(=\)
  Female.
\item
  \(a_i \in \{1, \dots, 5\}\) denotes the age group: \(1\) \(=\)
  \([15,45)\); \(2\) \(=\) \([45,65)\); \(3\) \(=\) \([65,75)\); \(4\)
  \(=\) \([75,85)\); and \(5\) \(=\) \([85,\dots)\).
\item
  \(d_i \in \{1, \dots, 4\}\) denotes the dwelling stratum of the
  sampling design with the same meaning as introduced above.
\item
  \(m_i \in \{1, \dots, 542\}\) denotes the municipality. From now on,
  we will refer to the small area as \(m\) instead of \(k\).
\end{itemize}

Let us consider Model \eqref{eq:model3}, which in this case study can be
formulated as: \begin{equation}
\logit (\gamma_j(\boldsymbol{x}_i)) = \kappa_{s_i, a_i, j} + \alpha_{d_i} + \theta_{m_i}, \ j = 1,\dots,J-1. \label{eq:modelcs}
\end{equation} In this case, we assume the sex and age-group covariates
(and their interaction) to have different effects on the cumulative
probabilities; that is, individuals with different values of these
covariates would have different, completely independent, cut Prior
distributions for \(\boldsymbol{\kappa}\) and \(\boldsymbol{\theta}\)
are those described in \protect\hyperlink{sec:priors}{Section 2.2.1}. In
this case study, we assume contiguity between municipalities as the
criterion for defining the adjacency matrix for \(\boldsymbol{\theta}\).
For the dwelling fixed effects, we assume improper uniform priors
\(\{p(\alpha_d)\}_{d = 2}^4 \propto 1\), with
\(\alpha_1 = -\sum_{d = 2}^4 \alpha_d\). In this manner,
\(\boldsymbol{\kappa}\) and \(\boldsymbol{\theta}\) may be interpreted
as the corresponding effects for an average dwelling stratum. The model
was implemented in the MCMC programming language \texttt{WinBUGS}
(\protect\hyperlink{ref-Lunn2000}{Lunn \emph{et al.}, 2000}). The
\texttt{WinBUGS} code and the \texttt{R} code used to run the entire
study can be found at points. The need for these different cut points
will become evident below.
(\url{https://github.com/bsmiguelangel/bayesian-modeling-of-spatial-ordinal-data-from-health-surveys}).
We ran five chains in parallel with \(6000\) iterations per chain. Of
these, the first \(1000\) iterations were discarded as burn-in, and the
resulting chains were thinned to retain one out of every \(25\)
iterations. Therefore, a total of \(1000\) simulations from the
posterior distribution were saved for each parameter of interest.
Convergence was assessed by means of the Gelman-Rubin statistic and the
effective sample size. Convergence was checked for all the model
parameters. Specifically, the Gelman-Rubin statistic was not greater
than \(1.10\) and the effective number of simulations was not lower than
\(100\) for any of the parameters.

\begin{figure}[!h]
\centering
\includegraphics[width = 17cm]{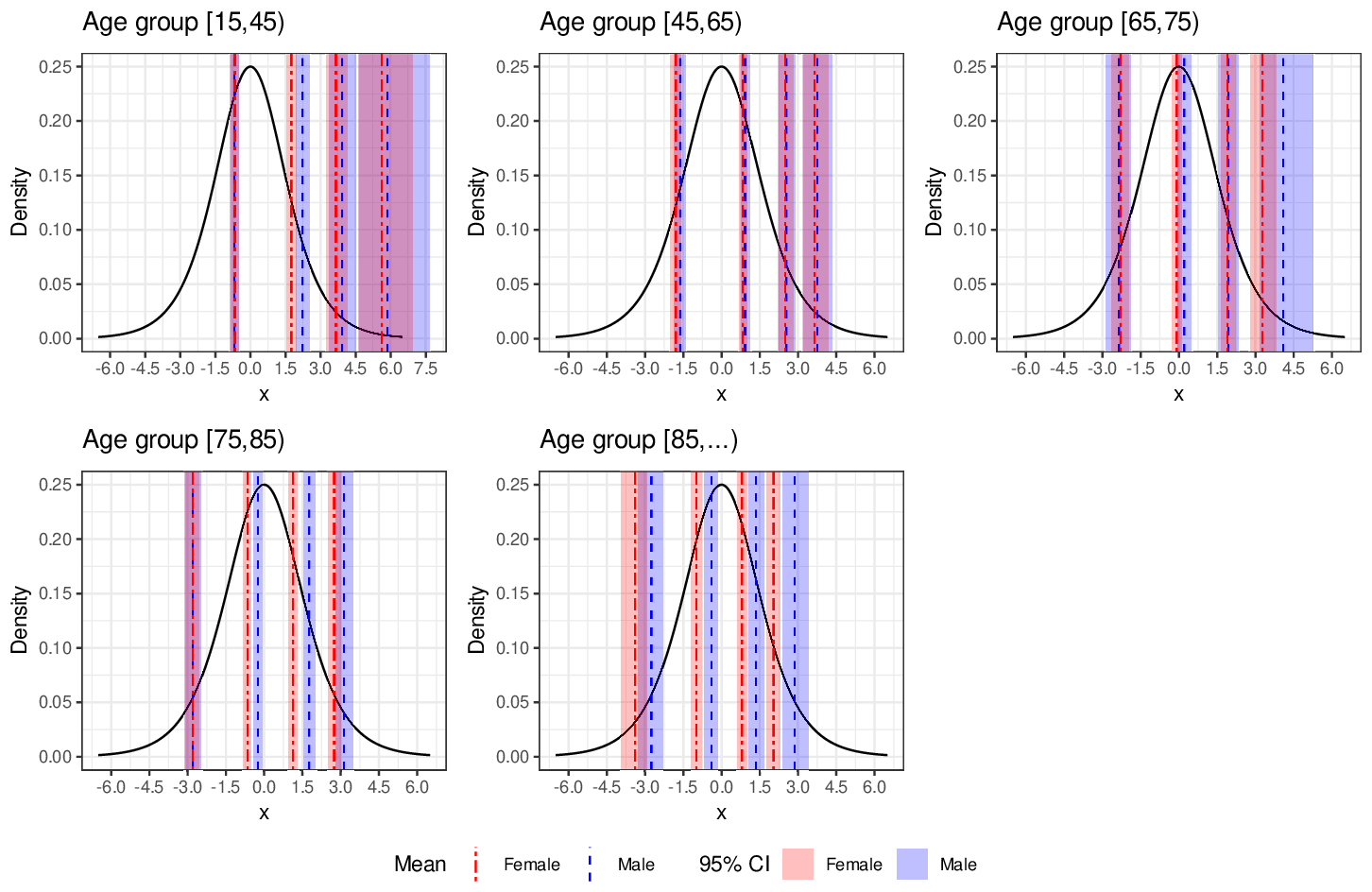}
\caption{Posterior distribution of all the cut points $\{\kappa_{saj}\}_{1\leq s \leq2, 1\leq a \leq 5, 1 \leq j \leq 4}$ associated with the five age group levels and sexes. Within each subfigure, both the posterior means and $95\%$ credible intervals of the cut points for each sex are shown.}
\label{fig:kappa}
\end{figure}

Figure \ref{fig:kappa} shows the standard logistic probability density
function and the posterior distribution of the cut points for each
combination of sex and age group levels, defining the probabilities of
each category as the area under the density function between two
corresponding consecutive cut points. According to this figure, we can
see that as age increases, the cut points move to the left; that is, a
higher probability of categories with worse self-perceived health is
estimated for the oldest groups. Moreover, interaction between sex and
age-group covariates is clearly necessary since, for example, sex has no
effect for the youngest groups but it has a clear effect for the oldest
ones. Specifically, young males and females seem to perceive similar
levels of health, while in the case of the oldest groups, females
clearly have lower cut points than males, pointing out worse
self-perceived health. In addition, interaction between those covariates
and the levels of the cumulative probabilities is also necessary, since
the cut points for males and females, within the same age group
(e.g.~\([75,85)\)), do not usually differ by an average fixed value.
Interestingly, note that \(95\%\) credible intervals of the last cut
point for both males and females within the age group \([15,45)\) are
the widest. This is possibly due to the fact that there are only a few
respondents within that age group who report ``Very bad'' self-perceived
health.

\begin{figure}[!h]
\centering
\includegraphics[width = 11.5cm]{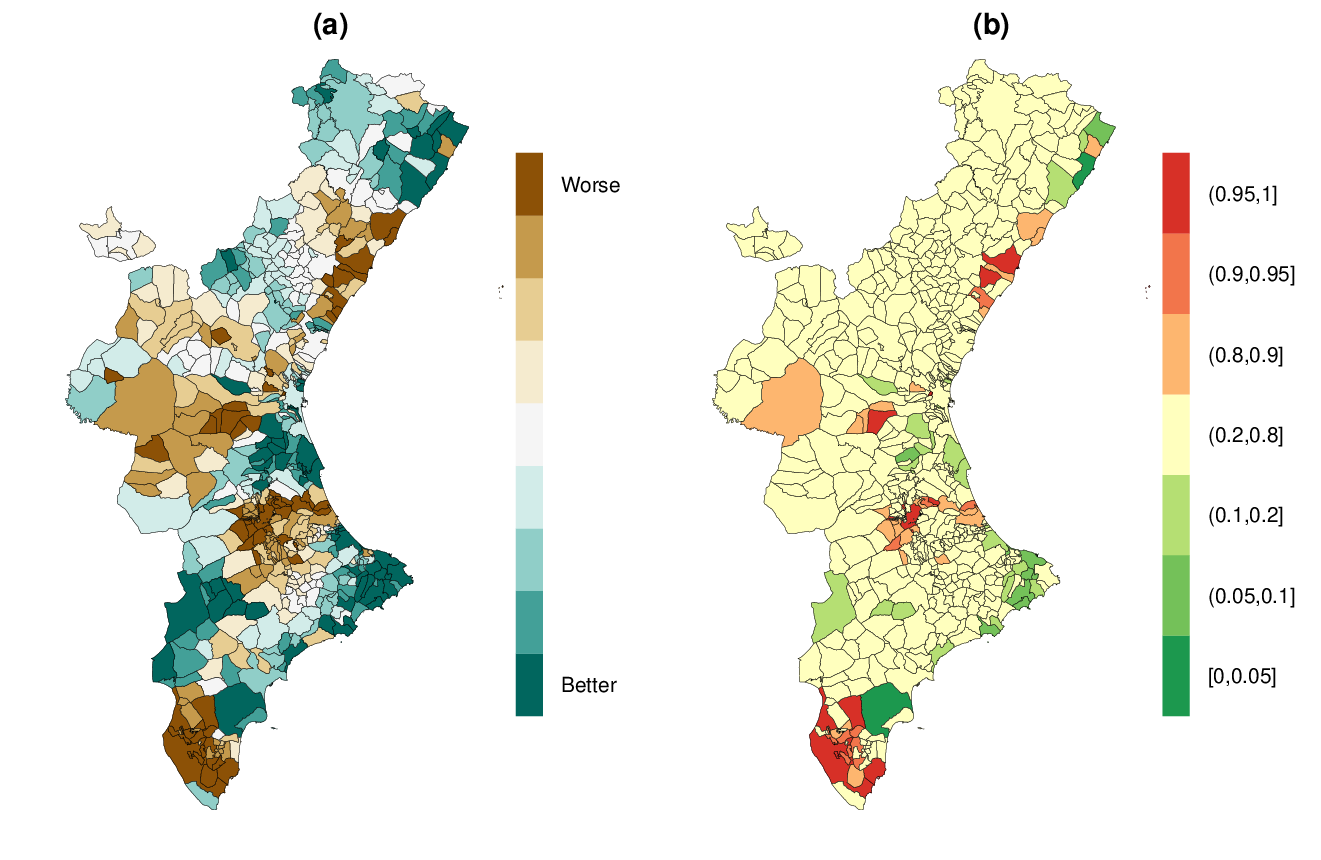}
\caption{(a) Posterior mean of the spatial random effect $\boldsymbol{\theta}$. (b) Relevance: $P(\theta_m < 0|\boldsymbol{y})$, $m = 1,\dots,542$.}
\label{fig:maps}
\end{figure}

Figure \ref{fig:maps} (a) shows the geographic pattern of the spatial
random effect, specifically the posterior means of the elements of
vector \(\boldsymbol{\theta}\). Each color refers to one of the nine
equal-probability intervals that have been used to discretize those
posterior means. This allows us to visualize which municipalities show
particularly worse or better self-perceived health. Indeed, the green
(brown) color represents the municipalities with better (worse)
self-perceived health. Note that \(\boldsymbol{\theta}\) accounts for
the geographic variability that is not explained by the covariates or,
in other words, it shows the remaining spatial pattern after controlling
for the effect of sex and age group on the available sample. Figure
\ref{fig:maps} (b) shows the relevance of the components of the spatial
random effect. Specifically, we measure this relevance as
\(P(\theta_m < 0|\boldsymbol{y})\) (posterior probability of worse
self-perceived health than the entire region as a whole),
\(m = 1,\dots,542\). Values of that probability close to one (zero)
imply a high (low) probability that the associated spatial random effect
is negative. Recall that negative values of the components of
\(\boldsymbol{\theta}\) make the cut points smaller (lower cumulative
probabilities), increasing so the probability of the last category
(``Very bad''). We consider in Figure \ref{fig:maps} (b) the spatial
effect to be relevant if that probability is greater than \(0.8\) or
less than \(0.2\). In summary, for those municipalities coloured in red
we have evidence of worse self-perceived health and for those in green
we find evidence of a better health.

It is important to emphasize here that the fixed effects of the dwelling
strata had no relevant association with the response variable (see
\protect\hyperlink{sec:supplementary}{Supplementary Information}); that is, it has no effect on self-perceived
health. Therefore, from now on, we assume its contribution linear
predictor term equal to zero. This assumption is equivalent to consider
no variability between dwelling strata, as suggested by the model
results.

\begin{figure}[!h]
\centering
\includegraphics[width = 18cm]{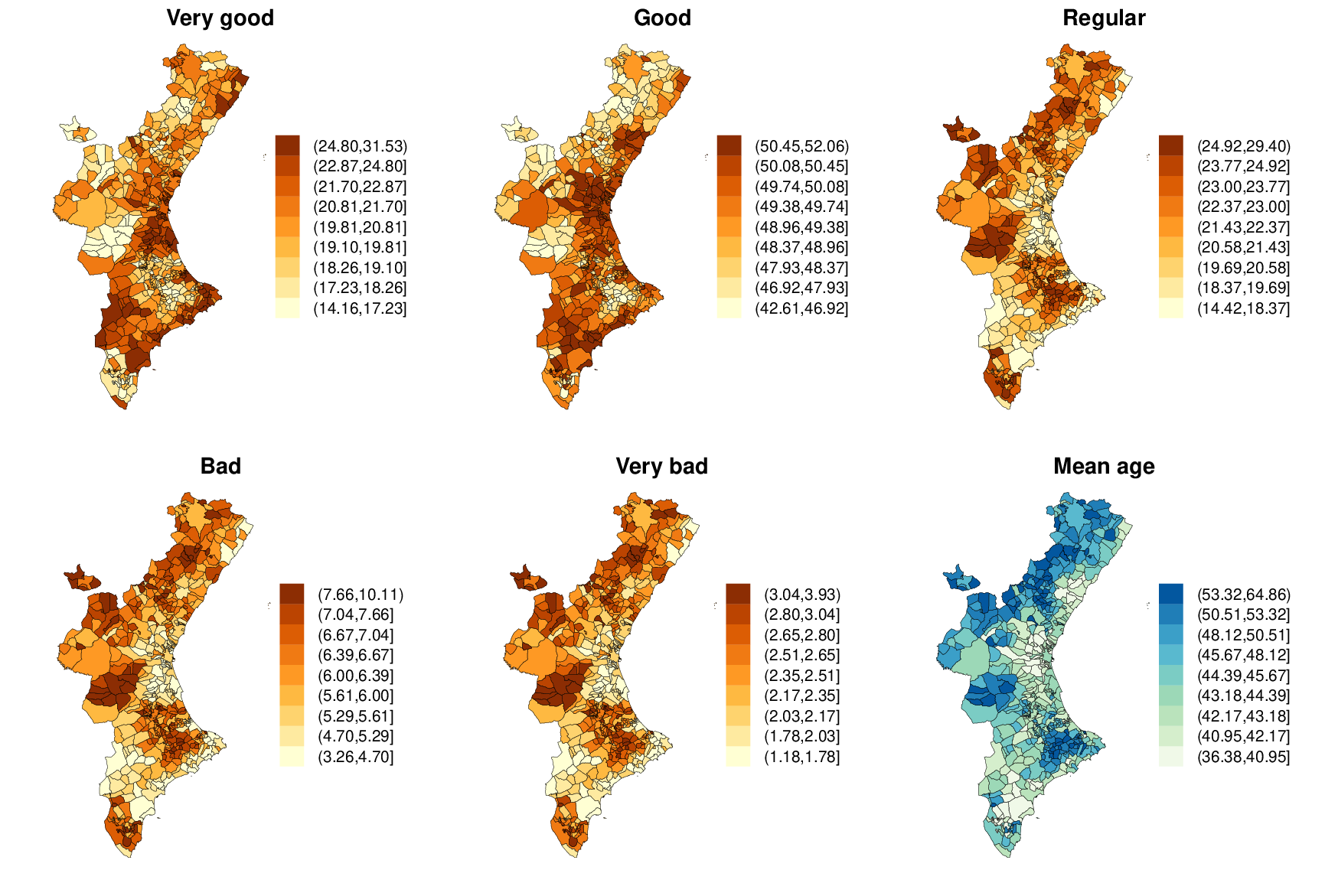}
\caption{Post-stratified posterior means of the percentages, at the area population level, for each category. The last map shows the mean age of each municipality.}
\label{fig:post}
\end{figure}

Figure \ref{fig:post} shows six maps. The first five maps show the
geographic pattern of population-level percentages for each of the
categories of the response variable; that is,
\(\hat{P}_{jm} \times 100\), \(j = 1,\dots,5\), \(m = 1,\dots,542\) (see
Expression \eqref{eq:post}). In a first stage, once the model has been
fitted, estimates of the probabilities for each sex \(s\), age group
\(a\) and municipality \(m\) from Model \eqref{eq:modelcs}
(\(\hat{\pi}_{jsam}\), \(j = 1,\dots,J\)) are obtained. In a second
stage, those estimates have been post-stratified using Expression
\eqref{eq:post} in order to compute the population estimates at the
municipality level. Note that the population of each sex \(s\), age
group \(a\) and municipality \(m\), \(N_{sam}\), must be known. Those
population totals have been downloaded from the National Institute of
Statistics
(\url{https://www.ine.es/dynt3/inebase/index.htm?padre=6225}). Unlike
the results in Figure \ref{fig:maps}, which represent model estimates
that summarize the sample information, the results in Figure
\ref{fig:post} are post-stratified to reproduce the population
composition of each municipality. Therefore, Figure \ref{fig:maps}
reproduces the self-perceived health controlling the effects of
individual covariates (age group, sex and dwelling strata), while Figure
\ref{fig:post} merges all the information for each municipality in order
to derive estimates which reproduce both the self-perceived health for
each population group and the population composition of each
municipality. It can be appreciated that the geographical distribution
of the categories corresponding to worse self-perceived health status
(``Regular'', ``Bad'' and ``Very bad'') resembles closely that of the
mean age (last map); that is, areas with worse self-perceived health
coincide with those with older population. Thus, Figure \ref{fig:maps}
is able to filter out the effect of the population composition that
Figure \ref{fig:post} does reproduce. The convenience of Figure
\ref{fig:maps} or Figure \ref{fig:post}, in practice, will depend on the
goal of the study (if decision makers want to reproduce or filter out
the effect of covariates in our estimates). In any case, the proposed
model allows both representations to be obtained, which is an important
benefit.

Finally, we assess the fit of Model \eqref{eq:modelcs} by simulating
from the posterior predictive distribution of each sex-age
group-municipality combination and aggregating the predicted values for
each municipality. As mentioned before, once the model has been fitted,
estimates of the probabilities for each sex \(s\), age group \(a\) and
municipality \(m\) (\(\hat{\pi}_{jsam}\), \(j = 1,\dots,J\)) are
obtained. Subsequently, we simulate \(n_{sam}\) (the sample size for
each sex, age group and municipality) values of the (categorical)
response variable from the corresponding posterior predictive
distribution. Finally, the estimated number of respondents in each
category is the result of adding all the previous results for each sex
and age group for each municipality. Specifically, Table \ref{tab:mc}
contains the corresponding posterior means, \(95\%\) prediction
intervals and the observed sample percentages for the four
municipalities with the largest population in the RV. As observed, there
are no large discrepancies between the observed and predicted values,
except for the first two categories in Elche. It seems that respondents
from that municipality report an excess of ``Very good'' self-perceived
health status at the expense of those who report merely ``Good'' health,
who are fewer than they should be according to the model.

\begin{table}[H]
\begin{center}
\resizebox{17cm}{!} {
\begin{tabular}{|c|ccccc|}
\hline
Municipality &
  \multicolumn{1}{c|}{Very good} &
  \multicolumn{1}{c|}{Good} &
  \multicolumn{1}{c|}{Regular} &
  \multicolumn{1}{c|}{Bad} &
  Very bad \\ \hline
Valencia &
  \multicolumn{1}{c|}{\begin{tabular}[c]{@{}c@{}} 13.02 (10.13-16.01) \\ \textbf{12.09} \end{tabular}} &
  \multicolumn{1}{c|}{\begin{tabular}[c]{@{}c@{}} 40.7 (36.44-45.26) \\ \textbf{42.97} \end{tabular}} &
  \multicolumn{1}{c|}{\begin{tabular}[c]{@{}c@{}} 31.31 (27.45-35.29) \\ \textbf{32.19} \end{tabular}} &
  \multicolumn{1}{c|}{\begin{tabular}[c]{@{}c@{}} 10.59 (7.84-13.24) \\ \textbf{7.35} \end{tabular}} &
  \begin{tabular}[c]{@{}c@{}} 4.38 (2.61-6.37) \\ \textbf{5.39} \end{tabular} \\ \hline
Alicante &
  \multicolumn{1}{c|}{\begin{tabular}[c]{@{}c@{}} 14.19 (10.06-18.05) \\ \textbf{12.13} \end{tabular}} &
  \multicolumn{1}{c|}{\begin{tabular}[c]{@{}c@{}} 41.37 (36.09-46.75) \\ \textbf{41.42} \end{tabular}} &
  \multicolumn{1}{c|}{\begin{tabular}[c]{@{}c@{}} 30.31 (25.15-35.5) \\ \textbf{34.62} \end{tabular}} &
  \multicolumn{1}{c|}{\begin{tabular}[c]{@{}c@{}} 10.01 (6.8-13.61) \\ \textbf{9.17} \end{tabular}} &
  \begin{tabular}[c]{@{}c@{}} 4.12 (2.07-6.51) \\ \textbf{2.66} \end{tabular} \\ \hline
Elche &
  \multicolumn{1}{c|}{\begin{tabular}[c]{@{}c@{}} 20.67 (15.87-26.05) \\ \textbf{29.34} \end{tabular}} &
  \multicolumn{1}{c|}{\begin{tabular}[c]{@{}c@{}} 45.77 (40.12-51.5) \\ \textbf{33.83} \end{tabular}} &
  \multicolumn{1}{c|}{\begin{tabular}[c]{@{}c@{}} 24.17 (19.16-29.04) \\ \textbf{27.84} \end{tabular}} &
  \multicolumn{1}{c|}{\begin{tabular}[c]{@{}c@{}} 6.82 (3.89-9.88) \\ \textbf{5.99} \end{tabular}} &
  \begin{tabular}[c]{@{}c@{}} 2.57 (0.9-4.5) \\ \textbf{2.99} \end{tabular} \\ \hline
\begin{tabular}[c]{@{}c@{}}Castellón\\ de la Plana\end{tabular} &
  \multicolumn{1}{c|}{\begin{tabular}[c]{@{}c@{}} 12.27 (7.03-17.84) \\ \textbf{14.05} \end{tabular}} &
  \multicolumn{1}{c|}{\begin{tabular}[c]{@{}c@{}} 40.5 (33.51-47.57) \\ \textbf{37.3} \end{tabular}} &
  \multicolumn{1}{c|}{\begin{tabular}[c]{@{}c@{}} 30.87 (23.78-37.84) \\ \textbf{29.19} \end{tabular}} &
  \multicolumn{1}{c|}{\begin{tabular}[c]{@{}c@{}} 11.43 (6.49-16.22) \\ \textbf{13.51} \end{tabular}} &
  \begin{tabular}[c]{@{}c@{}} 4.93 (1.62-8.11) \\ \textbf{5.95} \end{tabular} \\ \hline
\end{tabular}
}
     \caption{Model assessment. Posterior means, ($95\%$ prediction intervals) and observed values (in bold) of the percentage of respondents in each category for the four most populated municipalities in the RV.}
     \label{tab:mc}
\end{center}
\end{table}

\hypertarget{sec:conclusions}{%
\section{Conclusions}\label{sec:conclusions}}

Health surveys are alternative information sources for health
surveillance that allow the study of health issues of interest that
regular health databases generally do not allow. In this paper, we
propose a Bayesian hierarchical model for small-area estimation of
ordinal variables from health surveys. The main contribution of our
proposal is that it allows to analyze ordinal survey-based data at the
individual level, estimating the effect of any covariate associated with
the response variable. In addition, the proposed methodology
simultaneously accounts for spatial dependence and sampling design
information at a single stage, and estimates can be made at any desired
level of the model hierarchy via post-stratification. The representation
of the results with maps avoids the detailed, extensive and, on some
occasions, unassimilable information normally provided by tables in
health survey reports. Moreover, on the one hand, our model provides
population estimates of the cell frequencies of the outcome variable at
the small-area level. However, on the other hand, it also allows
estimating the geographic distribution of the outcome variable,
controlling for the effect of some specific covariates that it would be
possibly desirable to filter out their effect, such as age or sex. This
allows a deeper understanding and analysis of the data set, with
enhanced results and tools. These results could be used by public health
practitioners to intervene specifically in those areas of the region
with a higher risk than the rest.

As future lines of research, it would be interesting the development of
a methodology based on multivariate models, which would allow us to
jointly study sets of ordinal response variables likely to be
correlated. In fact, questions in health surveys are usually grouped in
blocks and these multivariate analyses would allow these variables to be
considered as true groups. The multivariate approach will help to
estimate the correlation between the different geographical patterns,
use that correlation in order to improve the estimation of those
geographical patterns and also control the individual-level effect.
Furthermore, it would also be interesting to apply this methodology to
national, instead of regional, health surveys at the municipality level,
even though it would be a challenge due to the expected high
computational cost. Finally, another line of research could be to extend
the current methodology to the spatio-temporal analysis of consecutive
editions of a common health survey in the same study region. As
mentioned above, it is common for health surveys to be collected
periodically (e.g.~every five years), so analyses of this kind could
visualize whether there are relevant changes over time.

\hypertarget{acknowledgments}{%
\section*{Acknowledgments}\label{acknowledgments}}
\addcontentsline{toc}{section}{Acknowledgments}

This study has been funded by Instituto de Salud Carlos III (ISCIII)
through the project ``PI19/00219'' and co-funded by the European Union.

\hypertarget{conflict-of-interest-statement}{%
\section*{Conflict of interest
statement}\label{conflict-of-interest-statement}}
\addcontentsline{toc}{section}{Conflict of interest statement}

The authors declare no conflicts of interest.

\hypertarget{data-availability-statement}{%
\section*{Data availability
statement}\label{data-availability-statement}}
\addcontentsline{toc}{section}{Data availability statement}

All the code used for the statistical analysis is shared as a
reproducible script at
(\url{https://github.com/bsmiguelangel/bayesian-modeling-of-spatial-ordinal-data-from-health-surveys}).
Unfortunately, due to the individual nature of the data set used, we are
not allowed to share it. However, data providers should allow access to
the data set under the same conditions as the authors of this study.

\hypertarget{references}{%
\section*{References}\label{references}}
\addcontentsline{toc}{section}{References}

\hypertarget{refs}{}
\begin{CSLReferences}{1}{0}
\leavevmode\vadjust pre{\hypertarget{ref-Binder1983}{}}%
\textsc{Binder}, D.A. (1983) \href{https://doi.org/10.2307/1402588}{On
the variances of asymptotically normal estimators from complex surveys}.
\emph{International Statistical Review}, \textbf{51}, 279--292.

\leavevmode\vadjust pre{\hypertarget{ref-Buttice2017}{}}%
\textsc{Buttice}, M.K. \& \textsc{Highton}, B. (2017)
\href{https://doi.org/10.1093/pan/mpt017}{How does multilevel regression
and poststratification perform with conventional national surveys?}
\emph{Political Analysis}, \textbf{21}, 449--467.

\leavevmode\vadjust pre{\hypertarget{ref-Chen2014}{}}%
\textsc{Chen}, C., \textsc{Wakefield}, J., \& \textsc{Lumely}, T. (2014)
\href{https://doi.org/10.1016/j.sste.2014.07.002}{The use of sampling
weights in bayesian hierarchical models for small area estimation}.
\emph{Spatial and Spatio-temporal Epidemiology}, \textbf{11}, 33--43.

\leavevmode\vadjust pre{\hypertarget{ref-Congdon2005}{}}%
\textsc{Congdon}, P. (2005)
\emph{\href{https://doi.org/10.1002/0470092394}{Bayesian models for
categorical data}}. John Wiley \& Sons, Ltd.

\leavevmode\vadjust pre{\hypertarget{ref-Congdon2022}{}}%
\textsc{Congdon}, P. \& \textsc{Lloyd}, P. (2022)
\href{https://doi.org/10.6339/JDS.2010.08(2).583}{Estimating small area
diabetes prevalence in the US using the behavioral risk factor
surveillance system}. \emph{Journal of Data Science}, \textbf{8},
235--252.

\leavevmode\vadjust pre{\hypertarget{ref-Desmee2015}{}}%
\textsc{Desmée}, S., \textsc{Mentré}, F., \textsc{Veyrat-Follet}, C., \&
\textsc{Guedj}, J. (2015)
\href{https://doi.org/10.1208/s12248-015-9745-5}{Nonlinear mixed-effect
models for prostate-specific antigen kinetics and link with survival in
the context of metastatic prostate cancer: A comparison by simulation of
two-stage and joint approaches}. \emph{The AAPS Journal}, \textbf{17},
691--699.

\leavevmode\vadjust pre{\hypertarget{ref-Dobson2018}{}}%
\textsc{Dobson}, A. \& \textsc{Barnett}, A. (2018)
\emph{\href{https://doi.org/10.1201/9781315182780}{An introduction to
generalized linear models}}, Fourth Edition (4th ed.) ed. Chapman \&
Hall/CRC.

\leavevmode\vadjust pre{\hypertarget{ref-Downes2020}{}}%
\textsc{Downes}, M. \& \textsc{Carlin}, J.B. (2020)
\href{https://doi.org/10.1093/aje/kwaa053}{Multilevel regression and
poststratification versus survey sample weighting for estimating
population quantities in large population health studies}.
\emph{American Journal of Epidemiology}, \textbf{189}, 717--725.

\leavevmode\vadjust pre{\hypertarget{ref-Downes2018}{}}%
\textsc{Downes}, M., \textsc{Gurrin}, L.C., \textsc{English}, D.R.,
\textsc{Pirkis}, J., \textsc{Currier}, D., \textsc{Spittal}, M.J., \&
\textsc{Carlin}, J.B. (2018)
\href{https://doi.org/10.1093/aje/kwy070}{Multilevel regression and
poststratification: A modeling approach to estimating population
quantities from highly selected survey samples}. \emph{American Journal
of Epidemiology}, \textbf{187}, 1780--1790.

\leavevmode\vadjust pre{\hypertarget{ref-Faraway2016}{}}%
\textsc{Faraway}, J. (2016)
\emph{\href{https://doi.org/10.1201/9781315382722}{Extending the linear
model with r: Generalized linear, mixed effects and nonparametric
regression models}}, Second Edition (2nd ed.) ed. Chapman \& Hall/CRC.

\leavevmode\vadjust pre{\hypertarget{ref-FH1979}{}}%
\textsc{Fay}, R.I. \& \textsc{Herriot}, R. (1979)
\href{https://doi.org/10.1080/01621459.1979.10482505}{Estimates of
income for small places: An application of james-stein procedures to
census data}. \emph{Journal of the American Statistical Association},
\textbf{74}, 269--277.

\leavevmode\vadjust pre{\hypertarget{ref-Gao2021}{}}%
\textsc{Gao}, Y., \textsc{Kennedy}, L., \textsc{Simpson}, D., \&
\textsc{Gelman}, A. (2021)
\href{https://doi.org/10.1214/20-BA1223}{Improving multilevel regression
and poststratification with structured priors}. \emph{Bayesian
Analysis}, \textbf{16}, 719--744.

\leavevmode\vadjust pre{\hypertarget{ref-Gelman2007}{}}%
\textsc{Gelman}, A. (2007)
\href{https://doi.org/10.1214/088342306000000691}{Struggles with survey
weighting and regression modeling}. \emph{Statistical Science},
\textbf{22}, 153--164.

\leavevmode\vadjust pre{\hypertarget{ref-Gelman2013}{}}%
\textsc{Gelman}, A., \textsc{Carlin}, J., \textsc{Stern}, H.,
\textsc{Dunson}, D., \textsc{Vehtari}, A., \& \textsc{Rubin}, D. (2013)
\emph{\href{https://doi.org/10.1201/b16018}{Bayesian data analysis}},
3rd ed. Chapman \& Hall/CRC.

\leavevmode\vadjust pre{\hypertarget{ref-Gelman2018}{}}%
\textsc{Gelman}, A., \textsc{Lax}, J., \textsc{Phillips}, J., \&
\textsc{Gabry}, J. (2018)
\href{http://www.stat.columbia.edu/~gelman/research/unpublished/MRT(1).pdf}{Using
multilevel regression and poststratification to estimate dynamic public
opinion}.

\leavevmode\vadjust pre{\hypertarget{ref-Gelman1997}{}}%
\textsc{Gelman}, A. \& \textsc{Little}, T. (1997)
\href{https://www150.statcan.gc.ca/n1/en/catalogue/12-001-X19970023616}{Poststratification
into many categories using hierarchical logistic regression}.
\emph{Survey Methodology}.

\leavevmode\vadjust pre{\hypertarget{ref-Hodges2010}{}}%
\textsc{Hodges}, J.S. \& \textsc{Reich}, B.J. (2010)
\href{https://doi.org/10.1198/tast.2010.10052}{Adding
spatially-correlated errors can mess up the fixed effect you love}.
\emph{The American Statistician}, \textbf{64}, 325--334.

\leavevmode\vadjust pre{\hypertarget{ref-HT1952}{}}%
\textsc{Horvitz}, D. \& \textsc{Thompson}, D. (1952)
\href{https://doi.org/10.1080/01621459.1952.10483446}{A generalization
of sampling without replacement from a finite universe}. \emph{Journal
of the American Statistical Association}, \textbf{47}, 663--685.

\leavevmode\vadjust pre{\hypertarget{ref-Lawson2008}{}}%
\textsc{Lawson}, A.B. (2008)
\emph{\href{https://doi.org/10.1201/9781584888413\%20}{Bayesian disease
mapping: Hierarchical modeling in spatial epidemiology}}. Chapman \&
Hall/CRC.

\leavevmode\vadjust pre{\hypertarget{ref-Leroux2000}{}}%
\textsc{Leroux}, B.G., \textsc{Lei}, X., \& \textsc{Breslow}, N. (2000)
Estimation of disease rates in small areas: A new mixed model for
spatial dependence. \emph{Statistical Models in Epidemiology, the
Environment, and Clinical Trials}, DOI:
\href{https://doi.org/10.1007/978-1-4612-1284-3_4}{10.1007/978-1-4612-1284-3\_4}.

\leavevmode\vadjust pre{\hypertarget{ref-Little1993}{}}%
\textsc{Little}, R. (1993)
\href{https://doi.org/10.1080/01621459.1993.10476368}{Post-stratification:
A modeler's perspective}. \emph{Journal of the American Statistical
Association}, \textbf{88}, 1001--1012.

\leavevmode\vadjust pre{\hypertarget{ref-Little2003}{}}%
\textsc{Little}, R.J. (2003)
\href{https://doi.org/10.1002/0470867205.ch4}{The bayesian approach to
sample survey inference}, 49--57.

\leavevmode\vadjust pre{\hypertarget{ref-Lunn2000}{}}%
\textsc{Lunn}, D.J., \textsc{Thomas}, A., \textsc{Best}, N., \&
\textsc{Spiegelhalter}, D. (2000)
\href{https://doi.org/10.1023/A:1008929526011}{WinBUGS -- a bayesian
modelling framework: Concepts, structure, and extensibility}.
\emph{Statistics and Computing}, \textbf{10}, 325--337.

\leavevmode\vadjust pre{\hypertarget{ref-Marhuenda2013}{}}%
\textsc{Marhuenda}, Y., \textsc{Molina}, I., \& \textsc{Morales}, D.
(2013) \href{https://doi.org/10.1016/j.csda.2012.09.002}{Small area
estimation with spatio-temporal fay-herriot models}. \emph{Computational
Statistics and Data Analysis}, \textbf{58}, 308--325.

\leavevmode\vadjust pre{\hypertarget{ref-Mari2014}{}}%
\textsc{Marí-Dell'Olmo}, M., \textsc{Martinez-Beneito}, M.,
\textsc{Gotsens}, M., \& \textsc{Palència}, L. (2014)
\href{https://doi.org/10.1007/s00477-013-0782-2}{A smoothed ANOVA model
for multivariate ecological regression}. \emph{Stochastic Environmental
Research and Risk Assessment}, 695--706.

\leavevmode\vadjust pre{\hypertarget{ref-Migue2019}{}}%
\textsc{Martínez-Beneito}, M. \& \textsc{Botella-Rocamora}, P. (2019)
\emph{\href{https://doi.org/10.1201/9781315118741}{Disease mapping: From
foundations to multidimensional modeling}}. Chapman \& Hall/CRC.

\leavevmode\vadjust pre{\hypertarget{ref-Mauff2020}{}}%
\textsc{Mauff}, K., \textsc{Steyerberg}, E., \textsc{Kardys}, I.,
\textsc{Boersma}, E., \& \textsc{Rizopoulos}, D. (2020)
\href{https://doi.org/10.1007/s11222-020-09927-9}{Joint models with
multiple longitudinal outcomes and a time-to-event outcome: A corrected
two-stage approach}. \emph{Statistics and Computing}, \textbf{30},
999--1014.

\leavevmode\vadjust pre{\hypertarget{ref-McCull1980}{}}%
\textsc{McCullagh}, P. (1980)
\href{http://www.jstor.org/stable/2984952}{Regression models for ordinal
data}. \emph{Journal of the Royal Statistical Society. Series B
(Methodological)}, \textbf{42}, 109--142.

\leavevmode\vadjust pre{\hypertarget{ref-Mercer2014}{}}%
\textsc{Mercer}, L., \textsc{Wakefield}, J., \textsc{Chen}, C., \&
\textsc{Lumley}, T. (2014)
\href{https://doi.org/10.1016/j.spasta.2013.12.001}{A comparison of
spatial smoothing methods for small area estimation with sampling
weights}. \emph{Spatial Statistics}, \textbf{8}, 69--85.

\leavevmode\vadjust pre{\hypertarget{ref-Mercer2015}{}}%
\textsc{Mercer}, L., \textsc{Wakefield}, J., \textsc{Pantazis}, A.,
\textsc{Lutambi}, A., \textsc{Masanja}, H., \& \textsc{Clark}, S. (2015)
\href{https://doi.org/10.1214/15-AOAS872}{Space--time smoothing of
complex survey data: Small area estimation for child mortality}.
\emph{Annals of Applied Statistics}, \textbf{9}, 1889--1905.

\leavevmode\vadjust pre{\hypertarget{ref-Paige2020}{}}%
\textsc{Paige}, J., \textsc{Fuglstad}, G.-A., \textsc{Riebler}, A., \&
\textsc{Wakefield}, J. (2020)
\href{https://doi.org/10.1093/jssam/smaa011}{Design- and model-based
approaches to small-area estimation in a low- and middle-income country
context: Comparisons and recommendations}. \emph{Journal of Survey
Statistics and Methodology}, \textbf{10}, 50--80.

\leavevmode\vadjust pre{\hypertarget{ref-Paisley2010}{}}%
\textsc{Paisley}, J. (2010)
\href{https://sciencedocbox.com/Physics/79278633-A-simple-proof-of-the-stick-breaking-construction-of-the-dirichlet-process.html}{A
simple proof of the stick-breaking construction of the dirichlet
process}. \emph{Technical report, Princeton University, Department of
Computer Science}.

\leavevmode\vadjust pre{\hypertarget{ref-Park2004}{}}%
\textsc{Park}, D.K., \textsc{Gelman}, A., \& \textsc{Bafumi}, J. (2004)
\href{https://doi.org/10.1093/pan/mph024}{Bayesian multilevel estimation
with poststratification: State-level estimates from national polls}.
\emph{Political Analysis}, \textbf{12}, 375--385.

\leavevmode\vadjust pre{\hypertarget{ref-Park2006}{}}%
\textsc{Park}, D.K., \textsc{Gelman}, A., \& \textsc{Bafumi}, J. (2006)
State-level opinions from national surveys: Poststratification using
multilevel logistic regression, DOI:
\href{https://doi.org/10.11126/stanford/9780804753005.003.0011}{10.11126/stanford/9780804753005.003.0011}.

\leavevmode\vadjust pre{\hypertarget{ref-Parker2020}{}}%
\textsc{Parker}, P.A., \textsc{Holan}, S.H., \& \textsc{Janicki}, R.
(2020) \href{https://doi.org/10.1002/sta4.267}{Conjugate bayesian
unit-level modelling of count data under informative sampling designs}.
\emph{Stat}, \textbf{9}, e267.

\leavevmode\vadjust pre{\hypertarget{ref-Parker2022}{}}%
\textsc{Parker}, P.A., \textsc{Holan}, S., \& \textsc{Janicki}, R.
(2022) \href{https://doi.org/10.1214/21-AOAS1524}{Computationally
efficient bayesian unit-level models for non-gaussian data under
informative sampling with application to estimation of health insurance
coverage}. \emph{The Annals of Applied Statistics}, \textbf{16},
887--904.

\leavevmode\vadjust pre{\hypertarget{ref-Peterson1990}{}}%
\textsc{Peterson}, B. \& \textsc{Harrell}, F.E. (1990)
\href{https://doi.org/10.2307/2347760}{Partial proportional odds models
for ordinal response variables}. \emph{Journal of the Royal Statistical
Society. Series C (Applied Statistics)}, \textbf{39}, 205--217.

\leavevmode\vadjust pre{\hypertarget{ref-Porter2014}{}}%
\textsc{Porter}, A., \textsc{Holan}, S., \textsc{Wikle}, C., \&
\textsc{Cressie}, N. (2014)
\href{https://doi.org/10.1016/j.spasta.2014.07.001}{Spatial fay-herriot
models for small area estimation with functional covariates}.
\emph{Spatial Statistics}, \textbf{10}, 27--42.

\leavevmode\vadjust pre{\hypertarget{ref-Reich2006}{}}%
\textsc{Reich}, B.J., \textsc{Hodges}, J.S., \& \textsc{Zadnik}, V.
(2006) \href{https://doi.org/10.1111/j.1541-0420.2006.00617.x}{Effects
of residual smoothing on the posterior of the fixed effects in
disease-mapping models}. \emph{Biometrics}, \textbf{62}, 1197--1206.

\leavevmode\vadjust pre{\hypertarget{ref-Sethuraman1994}{}}%
\textsc{Sethuraman}, J. (1994)
\href{http://www.jstor.org/stable/24305538}{A constructive definition of
dirichlet priors}. \emph{Statistica Sinica}, \textbf{4}, 639--650.

\leavevmode\vadjust pre{\hypertarget{ref-Skinner1989}{}}%
\textsc{Skinner}, C. (1989)
\emph{\href{https://eprints.soton.ac.uk/34696/}{Domain means, regression
and multi-variate analysis}}. Wiley.

\leavevmode\vadjust pre{\hypertarget{ref-Sun2022}{}}%
\textsc{Sun}, A., \textsc{Parker}, P.A., \& \textsc{Holan}, S.H. (2022)
\href{https://doi.org/10.3390/stats5010010}{Analysis of household pulse
survey public-use microdata via unit-level models for informative
sampling}. \emph{Stats}, \textbf{5}, 139--153.

\leavevmode\vadjust pre{\hypertarget{ref-Thompson2022}{}}%
\textsc{Thompson}, M.E., \textsc{Meng}, G., \textsc{Sedransk}, J.,
\textsc{Chen}, Q., \& \textsc{Anthopolos}, R. (2022)
\href{https://doi.org/10.1007/978-3-031-08329-7_13}{Spatial multilevel
modelling in the galveston bay recovery study survey}, 275--293.

\leavevmode\vadjust pre{\hypertarget{ref-Vandendijck2016}{}}%
\textsc{Vandendijck}, Y., \textsc{Faes}, C., \textsc{Kirby}, R.,
\textsc{Lawson}, A., \& \textsc{Hens}, N. (2016)
\href{https://doi.org/10.1016/j.spasta.2016.09.004}{Model-based
inference for small area estimation with sampling weights}.
\emph{Spatial Statistics}, \textbf{18}, 455--473.

\leavevmode\vadjust pre{\hypertarget{ref-Vergara2023}{}}%
\textsc{Vergara-Hernández}, C., \textsc{Marí-Dell'Olmo}, M.,
\textsc{Oliveras}, L., \& \textsc{Martínez-Beneito}, M. (2023) Taking
advantage of sampling designs in spatial small area survey studies, DOI:
\href{https://doi.org/10.48550/arXiv.2112.05468}{10.48550/arXiv.2112.05468}.

\leavevmode\vadjust pre{\hypertarget{ref-Wang2015}{}}%
\textsc{Wang}, W., \textsc{Rothschild}, D., \textsc{Goel}, S., \&
\textsc{Gelman}, A. (2015)
\href{https://doi.org/10.1016/j.ijforecast.2014.06.001}{Forecasting
elections with non-representative polls}. \emph{International Journal of
Forecasting}, \textbf{31}, 980--991.

\leavevmode\vadjust pre{\hypertarget{ref-Watjou2017}{}}%
\textsc{Watjou}, K., \textsc{Faes}, C., \textsc{Lawson}, A.,
\textsc{Kirby}, R., \textsc{Aregay}, M., \textsc{Carroll}, R., \&
\textsc{Vandendijck}, Y. (2017)
\href{https://doi.org/10.1002/sim.7369}{Spatial small area smoothing
models for handling survey data with nonresponse}. \emph{Statistics in
Medicine}, \textbf{36}, 3708--3745.

\leavevmode\vadjust pre{\hypertarget{ref-You2011}{}}%
\textsc{You}, Y. \& \textsc{Zhou}, Q. (2011)
\href{https://www150.statcan.gc.ca/n1/en/catalogue/12-001-X201100111445}{Hierarchical
bayes small area estimation under a spatial model with application to
health survey data}. \emph{Survey Methodology}, \textbf{37}, 25--37.

\leavevmode\vadjust pre{\hypertarget{ref-Zhang2018}{}}%
\textsc{Zhang}, X., \textsc{Holt}, J.B., \textsc{Lu}, H.,
\textsc{Wheaton}, A.G., \textsc{Ford}, E.S., \textsc{Greenlund}, K.J.,
\& \textsc{Croft}, J.B. (2014)
\href{https://doi.org/10.1093/aje/kwu018}{Multilevel regression and
poststratification for small-area estimation of population health
outcomes: A case study of chronic obstructive pulmonary disease
prevalence using the behavioral risk factor surveillance system}.
\emph{American Journal of Epidemiology}, \textbf{179}, 1025--1033.

\end{CSLReferences}

\newpage

\section*{Supplementary Information}\label{supplementary}
\addcontentsline{toc}{section}{Supplementary Information}

\begin{figure}[H]
\centering
\includegraphics[width = 16cm]{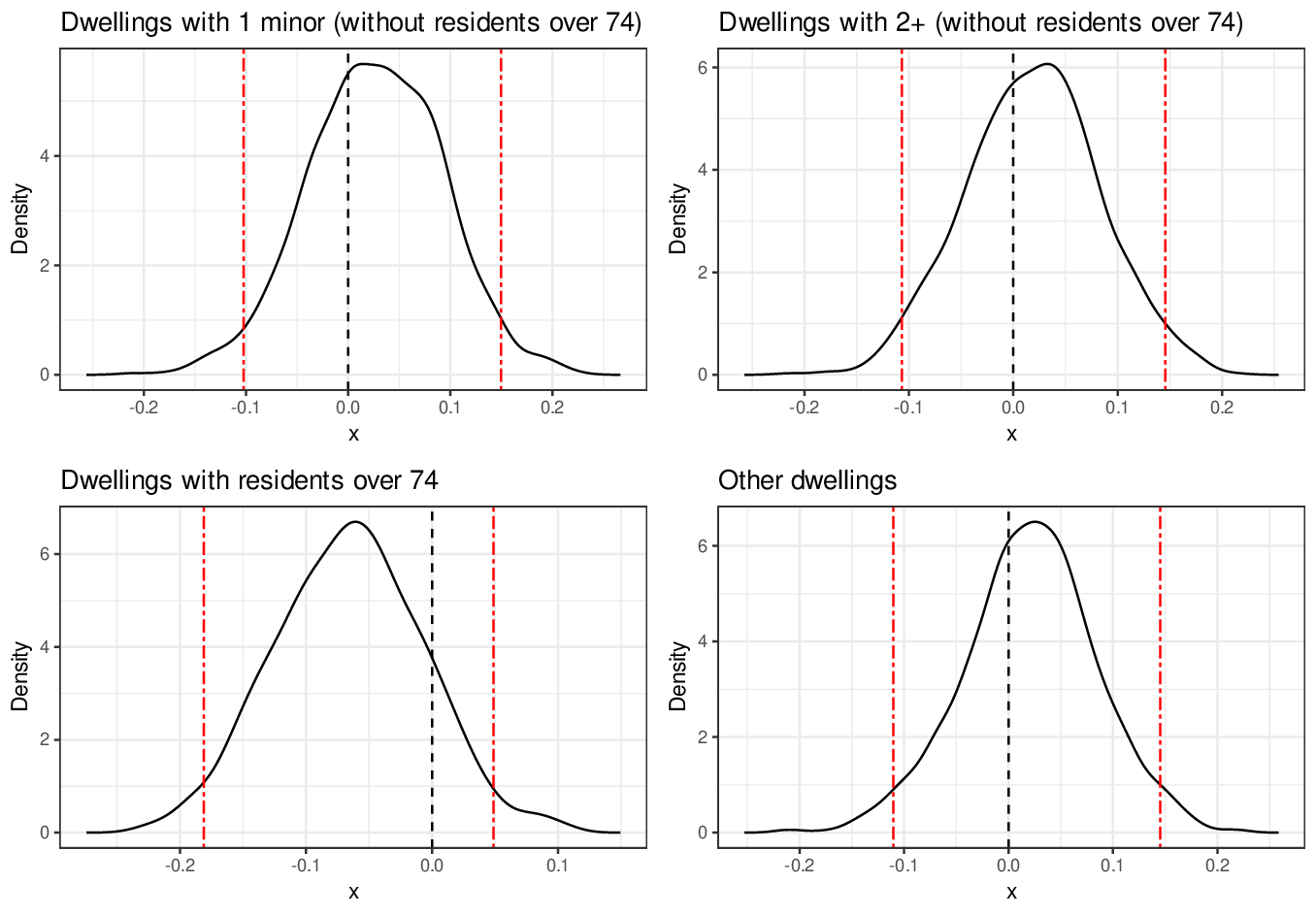}
\vspace{-0.4cm}
\caption{Posterior distribution of all the dwelling fixed effects. Within each subfigure, $95\%$ credible intervals are shown.}
\label{fig:alpha}
\end{figure}

\end{document}